\documentclass[twocolumn,showpacs,preprintnumbers,amsmath,amssymb]{revtex4}

\usepackage{graphicx}
\usepackage{bm}

\newcommand\Tm{\overline{\mathbf{T}}}
\newcommand\rr{{\bf r}}

\begin{document}

\title{Close or connected? Distance and connectivity effects on transport  in networks.}

\author{V. Tejedor}
\affiliation{Laboratoire de Physique Th\'eorique de la Mati\`ere Condens\'ee, UMR CNRS/UPMC, Universit\'e Pierre et Marie Curie, 4 Place Jussieu, 75255
Paris Cedex}

\author{O. B\'enichou}
\affiliation{Laboratoire de Physique Th\'eorique de la Mati\`ere Condens\'ee, UMR CNRS/UPMC, Universit\'e Pierre et Marie Curie, 4 Place Jussieu, 75255
Paris Cedex}

\author{R. Voituriez}
\affiliation{Laboratoire de Physique Th\'eorique de la Mati\`ere Condens\'ee, UMR CNRS/UPMC, Universit\'e Pierre et Marie Curie, 4 Place Jussieu, 75255
Paris Cedex}
\date{\today}

\begin{abstract}

We develop an analytical approach which provides the dependence of the mean first-passage time (MFPT) for random walks on complex networks both on the target connectivity and on the source-target distance.  Our  approach  puts forward 
two strongly different behaviors depending on the  type -- compact or non compact -- of the random walk. In the case of non compact exploration, we show that the MFPT  scales linearly with the inverse connectivity of the target,  and is largely independent of the starting point. On the contrary, in the compact case the MFPT is controlled by the source--target distance, and we find that unexpectedly the target connectivity becomes \textit{irrelevant} for remote targets.

\end{abstract}


\maketitle

\section{Introduction}
Complex networks theory is nowadays a common tool to analyze a broad class of phenomena in social, biological or physical sciences \cite{Albert:2002,Dorogovtsev:2008,Barratbook}.  An important issue in the field is to quantify the impact of  topological properties of a network on its transport properties. As a paradigm of transport process, random walks on complex networks have been intensely studied \cite{Noh:2004a,Bollt:2005a,nechaev2002,Samukhin:2008,Baronchelli:2008,Kitsak:2010fk}, and  the mean  first-passage time (MFPT) \cite{Redner:2001a}  to a target node  -- which quantifies the time needed for a random walker to find a target on the network -- has been widely used as an indicator of transport efficiency \cite{Kozak:2002,Huang:2006,Gallos:2007a,Haynes:2008,Agliari:2008,Agliari:2009,Zhang:2009,Zhang:2009a}.   

A striking topological feature of many real-world complex networks is the wide distribution of the number $k$ of links attached to a node -- the connectivity -- , as exemplified by the now celebrated class of   scale-free networks,  such as internet \cite{Faloutsos:99}, biological networks \cite{Noort:2004}, stock markets \cite{Kim:2002} or urban traffic \cite{Wu:2004},  for which the connectivity is distributed according to a power law. The impact of connectivity on transport properties has been put forward in  \cite{Lopez:2005fk, Kittas:2008,Baronchelli:2008,Tejedor:2009vn}, where it was found in different examples that transport towards a target node can be favored by  a high connectivity  of the  target, and different functional forms of the dependence of the MFPT on the target connectivity were proposed. On the other hand, the dependence of the MFPT on geometric properties, such as the volume of the network and the source to target distance, has been obtained recently in \cite{Condamin:2007zl,Benichou:2008,Condamin:2008,BenichouO.:2010a}, where it was shown  that the starting position of the random walker plays a crucial role in the target search problem. In this context, quantifying the relative importance of distance and connectivity effects on transport properties  on complex networks remains an important and widely unanswered question, which can be summarized as follows :  is it faster for a random walker to find either a close,  or a highly connected target?

Here, we propose a general framework, applicable to a broad class of networks, which deciphers the dependence of the MFPT both on the target connectivity and on the source to target distance,  and provides a global understanding of recent results obtained on specific examples. Our  approach  highlights 
two strongly different behaviors depending on the so--called type -- compact or non compact -- of the random walk. In the case of non compact exploration, the MFPT is found to scale linearly with the inverse connectivity of the target,  and to be widely independent of the starting point. On the contrary, in the compact case the MFPT is controlled by the source to target distance, and we find that unexpectedly the target connectivity is \textit{irrelevant} for remote targets.
 This analytical approach, validated numerically on various examples of networks, can be extended to other relevant first-passage observables such as splitting probabilities or occupations times \cite{Condamin:2008}.

\section{Model and notations}
We are interested in the MFPT denoted $\Tm({\bf r}_T|{\bf r}_S)$ of a discrete Markovian random walker to a target  ${\bf r}_T$, starting from a source point ${\bf r}_S$, and evolving in a network of $N$ sites. We denote by $k({\bf r})$ the connectivity (number of nearest neighbors) of site ${\bf r}$, and by $\langle k\rangle$ its average over all  sites with a flat measure. The corresponding degree distribution is denoted by $p(k)$. We assume that at each time step $n$, the walker, at site ${\bf r}$, jumps to one of the neighboring site with probability $1/k({\bf r})$. Let $P({\bf r},n|{\bf r}')$ be the propagator, i.e. the  probability 
that the walker is
at  ${\bf r}$ after $n$ steps, starting from  ${\bf r}'$.    The stationary probability distribution is then given by $P_{\rm stat}({\bf r})=k({\bf r})/N\langle k\rangle$, and it can be shown that detailed balance yields the following symmetry relation :
\begin{equation}\label{sym}
P({\bf r},n|{\bf r}')P_{\rm stat}({\bf r}')=P({\bf r}',n|{\bf r})P_{\rm stat}({\bf r}),
\end{equation}
which will prove to be useful.  

We consider networks with only short range degree correlations, namely such that $\left<k(\rr)k(\rr')\right>=\left<k\right>^2$ for $|\rr-\rr'|$ larger than a cut-off distance $R$, where the average is taken over all pairs $\rr,\rr'$ with $|\rr-\rr'|$ fixed. This hypothesis is verified in particular by networks whose Pearson assortativity coefficient \cite{Newman:2002} is $ 0$, such as Erdos-Renyi networks. It is however less restrictive since  local degree correlations can exist, and many networks actually comply with this assumption,  as exemplified below. The hypothesis of short range degree correlations implies in particular that the degree distribution in a shell of radius  $r>R$ is identical to the degree distribution $p(k)$ over the whole network, so that  
\begin{equation}
 \sum_{\rr'\backslash|\rr-\rr'|=r}P_{\rm stat}({\bf r'})\simeq N_\rr(r)/N
 \end{equation}
  where $ N_\rr(r)$ is the number of sites $\rr'$ such that $|\rr-\rr'|=r$.
We then introduce the weighted average at distance $r$ of a function $f$ of two space variables defined by 
\begin{equation}
\{ f({\bf r},{\bf r}')\}_{\bf r'}=\frac{N}{N_\rr(r)}\sum_{\rr'/|\rr-\rr'|=r}f({\bf r},{\bf r}') P_{\rm stat}({\bf r'}), 
 \end{equation}
 and the standard flat average 
\begin{equation}
\langle f({\bf r},{\bf r}') \rangle_{\bf r'}=\frac{1}{N_\rr(r)}\sum_{\rr'/|\rr-\rr'|=r}f({\bf r},{\bf r}') .
\end{equation}

\section{Scaling form of the propagator for scale invariant processes}
\begin{figure}[htb!]
\includegraphics[width =0.9\linewidth,clip]{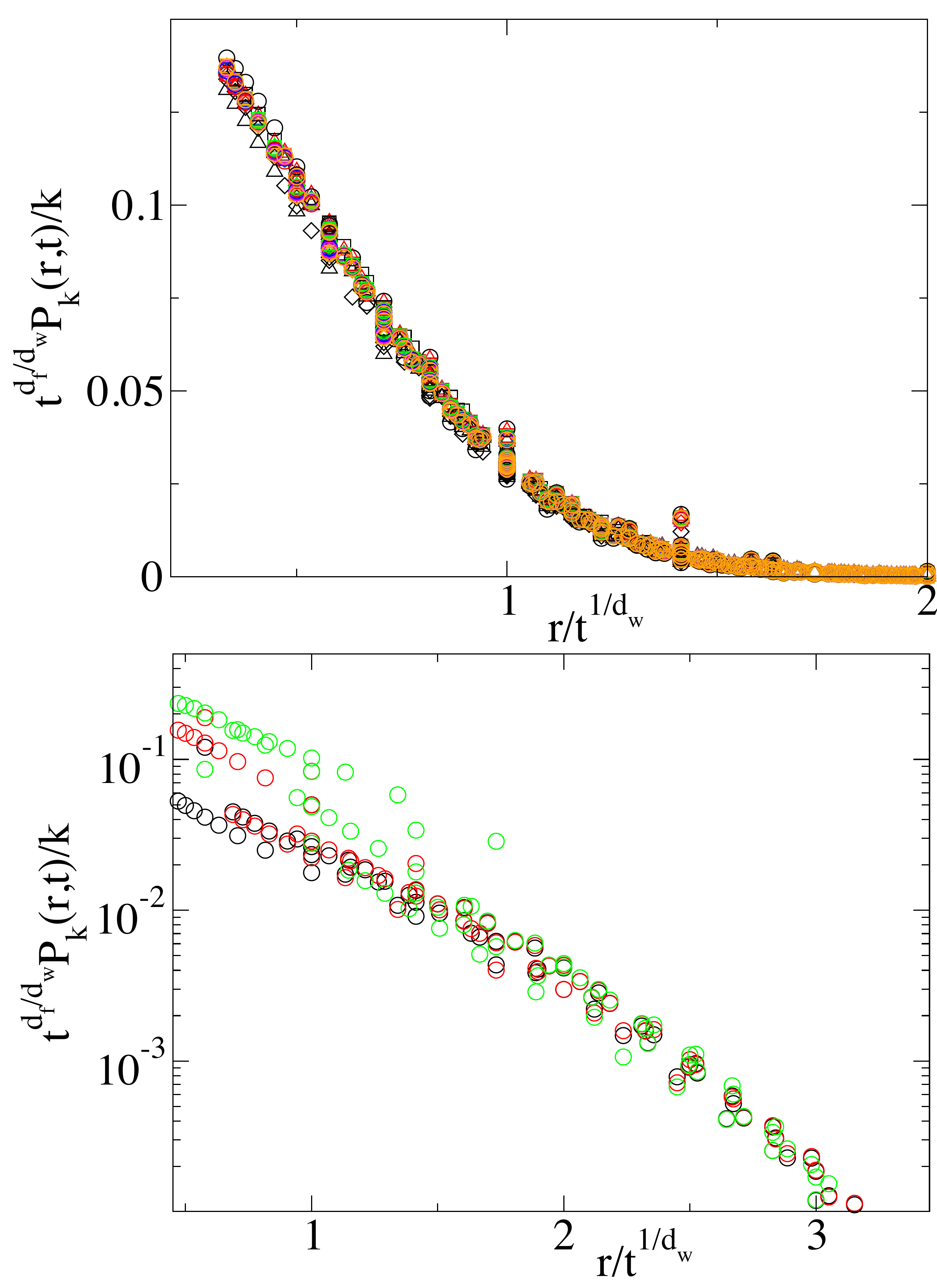}
\caption{Plot of the propagator $P(\rr_T,n|\rr_S)$ for non compact exploration. {\it Up:} Supercritical 3D percolation networks ($p=0.8$) of different sizes, and for different $k(\rr_T)$. $r_s$ is chosen in the center of the network, and $t$ is small enough to avoid hits on the network's border. Black, red, green, blue, magenta and orange symbols stand respectively for $k=1$, $2$, $3$, $4$, $5$ and $6$. Circles, triangles, diamonds and squares stand respectively for networks of size $20^3$, $25^3$, $30^3$ and $40^3$. {\it Down:} $(2,2,2)$-flowers. Black, red and green circles stand respectively for $k=2$, $6$ and $18$.\label{propnoncompact}}
\end{figure}

\begin{figure}[htb!]
\includegraphics[width =0.9\linewidth,clip]{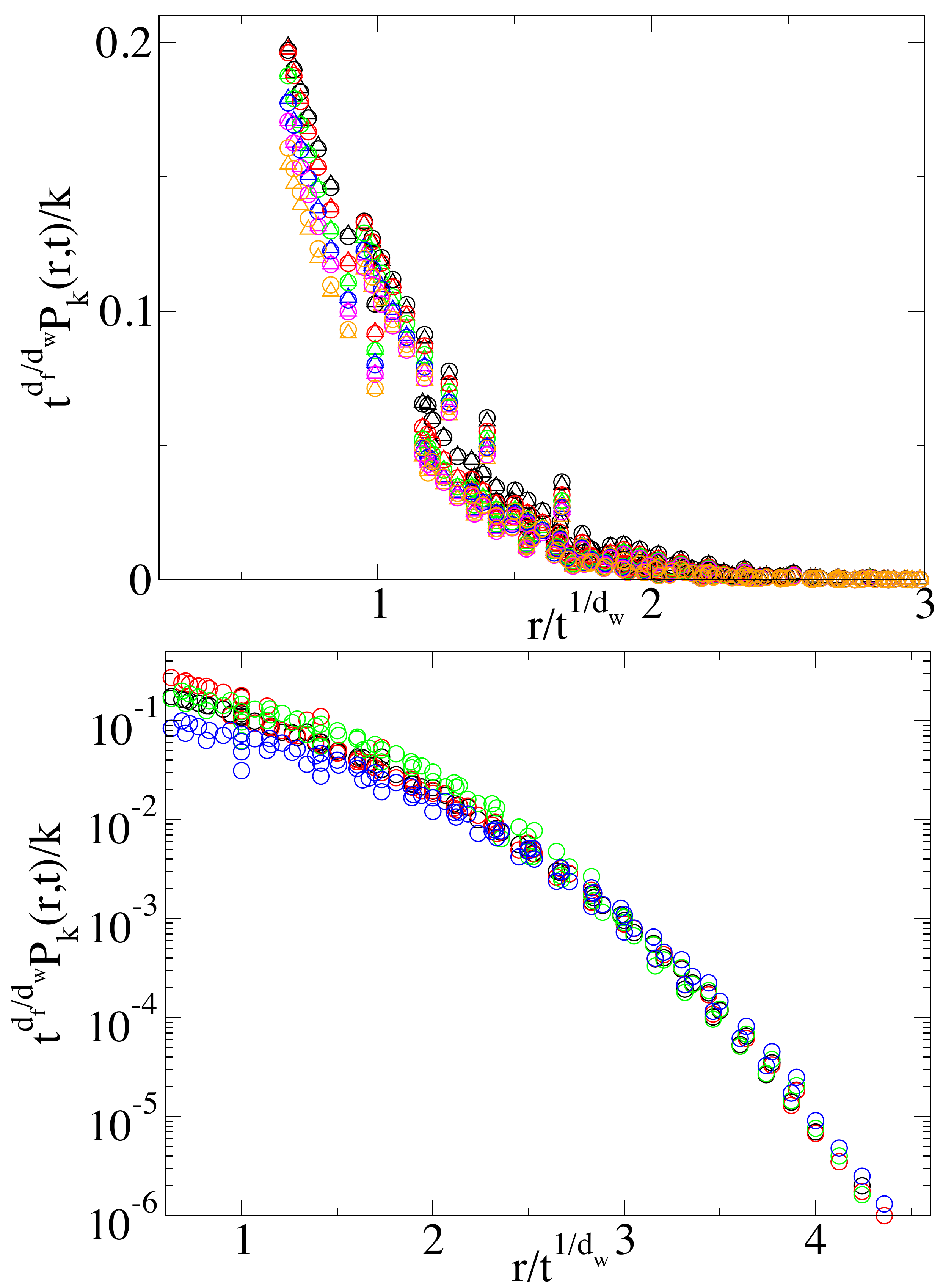}
\caption{Plot of the propagator $P(\rr_T,n|\rr_S)$ for compact exploration. {\it Up:} Critical 3D percolation networks ($p = 0.2488$) of different sizes, and for different $k(\rr_T)$. $r_s$ is chosen in the center of the network, and $t$ is small enough to avoid hits on the network's border. Black, red, green, blue, magenta and orange symbols stand respectively for $k=1$, $2$, $3$, $4$, $5$ and $6$. Circles, and triangles stand respectively for networks of size $40^3$ and $50^3$. {\it Down:} $(3,3)$-flowers (see \cite{Rozenfeld:2007}) for different $k(\rr_T)$. Black, red, green, blue circles stand respectively for $k=2$, $4$, $8$ and $16$.\label{propcompact}}
\end{figure}

We focus hereafter on  transport processes having  scale invariant properties. It this case,  we can assume that the propagator in the infinite network size limit   $P_0$, after averaging over points at a distance $r$ from the starting point, satisfies the standard scaling for  $|\rr-\rr'|>R$ :
\begin{equation}
\langle P_0({\bf r},n|{\bf r}')\rangle_{\rr} \propto  n^{-d_f/d_w} \Pi\left(
\frac{r}{n^{1/d_w}}\right),
\label{scaling}
\end{equation}
 where the fractal dimension $d_f$ characterizes the
accessible volume $V_r \propto r^{d_f}$ within a sphere of radius $r$,   and  the walk dimension $d_w$  characterizes  the distance $r \propto n^{1/d_w}$ covered by a random walker
in $n$ steps.
A first central result of this paper is to show numerically that the dependence of the propagator on the connectivity of the target site can be actually made explicit and reads
\begin{equation}
\langle P_0({\bf r},n|{\bf r}')\rangle_{\rr,k} \propto  k n^{-d_f/d_w} \Pi\left(\frac{r}{n^{1/d_w}}\right),
\label{scalingk}
\end{equation}
where the average is taken over sites $\rr$ at a distance $r$ from $\rr'$ with fixed connectivity $k$. An argument supporting the $k$ dependence hypothesized in  Eq. (\ref{scalingk}) is that it satisfies the symmetry relation of Eq. (\ref{sym}). Numerical simulations  on various examples of scale invariant networks, such as percolation clusters and $(u,v)$--flowers (see definition below) validate this assumption, as shown in Figs. \ref{propnoncompact} and \ref{propcompact}. We stress that the scaling form (\ref{scalingk}) is verified in the  cases  of  both compact ($d_w>d_f$) and non compact ($d_w<d_f$) exploration.  We believe that this result on its own can be important in the analysis of transport processes on networks. We show next that it enables to obtain the explicit dependence of first-passage properties on the connectivity of the target site.

\section{Mean first-passage time}
We now  extend the theory developed in \cite{Condamin:2007zl} to compute the MFPT of a discrete Markovian random walker to a target  ${\bf r}_T$, and obtain explicitly its dependence on $k(\rr_T)$. As shown in \cite{Noh:2004a,Condamin:2005db,Condamin:2007zl}, the MFPT satisfies the following exact expression :
\begin{equation}\label{Tm}
 P_{\rm stat}({\bf r}_T) \Tm({\bf r}_T|{\bf r}_S)= H({\bf r}_T|{\bf r}_T)-H({\bf r}_T|{\bf r}_S),
\end{equation}
where $H({\bf r}|{\bf r}') = \sum_{n=1}^{\infty} (P({\bf r},n|{\bf r}') -P_{\rm stat}({\bf r})  )$ is the  pseudo-Green function of the problem \cite{Barton:1989}.
Note that averaging equation (\ref{Tm}) for ${\bf r}_S$ covering the nearest neighbors of ${\bf r}_T$ gives the expression of the averaged  MFPT ${\langle  \Tm\rangle}_{\rm Kac}(\rr_T)$
expected from Kac formula \cite{Aldous:1999,Condamin:2007eu}:
\begin{equation}
 {\langle \Tm\rangle}_{\rm Kac}(\rr_T) =1/P_{\rm stat}(\rr_T)-1=N\langle k\rangle/k(\rr_T)-1,
\label{Kac}
\end{equation}
which we will use below.

Following \cite{Condamin:2007zl}, we consider the  large $N$ limit of Eq. (\ref{Tm}). Making use of Eq. (\ref{sym}), we obtain

\begin{equation}
P_{\rm stat}({\bf r}_T) \Tm({\bf r}_T|{\bf r}_S) \sim G_0({\bf r}_T|{\bf r}_T)-\frac{k({\bf r}_T)}{k({\bf r}_S)}G_0({\bf r}_S|{\bf r}_T).
\label{Tmleading}
\end{equation}
Here $G_{0}$ is the usual infinite space Green function defined by $G_0({\bf r}|{\bf r}') = \sum_{n=1}^\infty P_0({\bf r},n|{\bf r}') $, 
and   $\sim$ denotes equivalence for large $N$. It is useful to notice that this leading term of the MFPT still satisfies the Kac formula (\ref{Kac}).
We next take the weighted  average of Eq. (\ref{Tmleading}) over the source points and obtain:
\begin{equation}\label{Tmleading2}
 P_{\rm stat}({\bf r}_T)\Tm_{\rr_T}(r) \sim G_0({\bf r}_T|{\bf r}_T)-\frac{k({\bf r}_T)}{\left<k\right>}\langle G_0({\bf r}_S|{\bf r}_T)\rangle_{{\bf r}_S},
\end{equation}
where we defined  $\Tm_{\rr_T}(r)\equiv\{\Tm({\bf r}_T|{\bf r}_S)\}_{{\bf r}_S}$.
Substituting  the scaling (\ref{scaling}) in Eq. (\ref{Tmleading2}) then yields the large $N$ equivalence of the MFPT to a target site $\rr_T$ averaged over sources, which is valid for $r>R$:
\begin{equation}
\Tm_{\rr_T}(r) \sim N\langle k\rangle\left(
A_{k}+B r^{d_w-d_f} \label{scalingt}\right).
\end{equation}
In this expression the constant $A_{k}$ depends on the connectivity $k$ of the target and $B$ is a constant independent of $k$ ad $r$,  which depends on the scaling function $\Pi$. 
We now distinguish two regimes depending on the compact or non compact nature of the transport process, and focus on the large $r$ regime. 

\subsection{Compact case $d_w \geq d_f$}
In the compact  case, $d_w \geq d_f$, which corresponds to recurrent random walks, we obtain that the MFPT scales in the large $r$ limit as 
\begin{equation}\label{compactlarger}
\Tm_{\rr_T}(r) \sim N\langle k\rangle B r^{d_w-d_f}.
\end{equation}
This shows that unexpectedly the MFPT is asymptotically independent of the connectivity of the target, while the dependence on the distance $r$ is crucial.
Eq. (\ref{scalingt}) is valid for $r$ large enough (typically $r>R$). The dependence of $A_k$ on $k$, which impacts on the MFPT for $r$ small only,  can be estimated by  assuming that this expression still holds approximately for short distances.  Following \cite{Benichou:2008a}, we take $r=1$ in Eq. (\ref{scalingt}) and use the Kac formula (\ref{Kac}) to obtain:
\begin{equation}
1/k \approx 
 A_k+ B,
\label{scalingKac}
\end{equation}
which provides the $k$-dependence of $A_k$. We next aim at evaluating $B$. We introduce  the weighted average  of the MFPT over the target point $\tau(r)=\sum_{\rr_T} P_{\rm stat}({\bf r}_T)\Tm_{\rr_T}(r)$. Using Eq. (\ref{scalingKac}), this quantity writes:
\begin{equation}
\tau(r) \sim N\left(1+B\left<k\right> (r^{d_w-d_f}-1)\right).
\label{scaling3}
\end{equation}
In the case of compact exploration, the continuous space limit can be defined (see \cite{Benichou:2008a}) and imposes $\tau(r\to0)=0$. This extra equation, based on the existence of a continuous limit, enables to evaluate $B$ as  $B=1/\langle k \rangle$. Note that for fractal trees ($d_w-d_f = 1$) we recover the exact result $\tau(r)= Nr$. Finally one has : 
\begin{equation}
\Tm_{\rr_T}(r) \sim N\langle k\rangle \left(
 \frac{1}{k}+\frac{1}{\langle k \rangle} \left (r^{d_w-d_f} - 1 \right ) \right),
\end{equation}
which fully elucidates the dependence of the MFPT on $k$ and $r$. We recall here that this expression is originally derived for $r$ large, and that the small $r$ regime relies on the less controlled assumption that the scaling form of the propagator (\ref{scalingk}) holds for any distance $r$, and in particular that a continuous limit exists. It will however prove numerically to be accurate in various examples for all $r$ values.

\subsection{Non compact case $d_w < d_f$}
\begin{figure}[htb!]
\includegraphics[width =0.9\linewidth,clip]{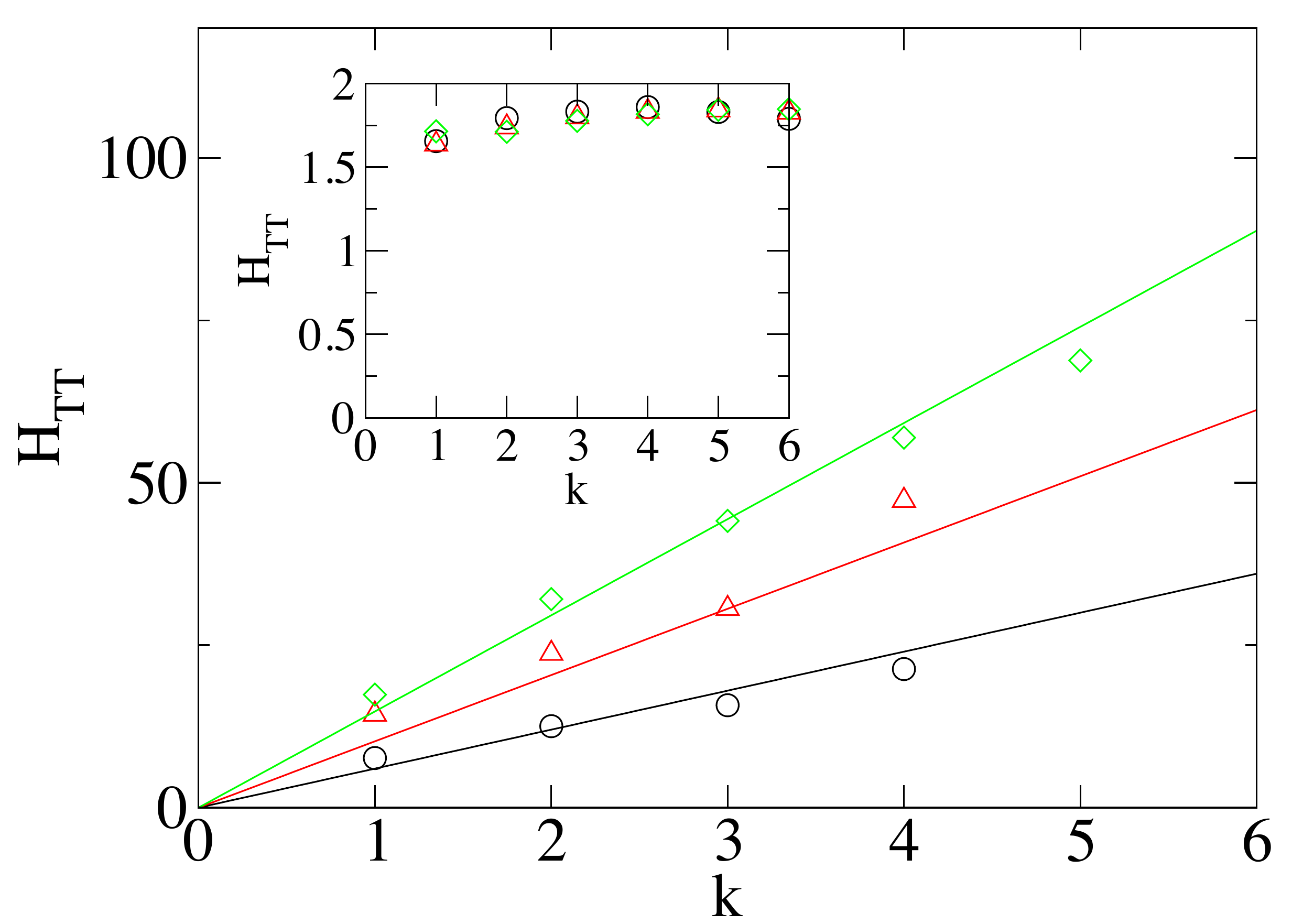}
\caption{Numerical computation of $H(\rr_T|\rr_T) $ averaged over a network of a given size, as a function of the target connectivity $k$, on supercritical ($p=0.8$) and critical ($p=0.2488$) 3D percolation network. The inset stands for the supercritical percolation network, for three sizes $10^3$ (black circles), $15^3$ (red triangles) and $20^3$ (green diamonds). Equation (\ref{scalinghtt}) gives $H(\rr_T|\rr_T)  = C$. The main figure stands for the critical percolation network, also for three sizes (same symbols), and a fit in $CkN^{d_w/d_f-1}$ (straight line).}
\label{fig:HTT}
\end{figure}

In the  non compact (or transient) case,  $d_w < d_f$, we obtain that the MFPT scales in the large $r$ limit as 
\begin{equation}
\Tm_{\rr_T}(r) \sim N\langle k\rangle A_k.
\label{scalingnoncompact}
\end{equation}
This shows that the MFPT is independent of $r$ for $r$ large, as was already discussed in the literature \cite{Condamin:2007zl}. The dependence on $k$ is now fully contained in the constant $A_k$, which we now determine. 
Following \cite{Kittas:2008}, we assume that the FPT distribution is proportional to $\exp ( -A k t/(N \langle k \rangle ))$, with $A = \mathcal{O}(1)$, and widely independent of $r$ in agreement with the result obtained in Eq.(\ref{scalingnoncompact}) for the first moment. This implies that the global MFPT, defined as the MFPT averaged over all source points and  denoted by $\{\Tm_{\rr_T}\}$,  scales as  $\{\Tm_{\rr_T}\} \propto N\langle k \rangle/k$. Using the exact result derived in \cite{Tejedor:2009vn} :
\begin{equation}
\{\Tm_{\rr_T} \}= \frac{ H({\bf r}_T|{\bf r}_T)}{P_{\rm stat}(\rr_T)}
\label{global}
\end{equation}
we obtain that $H({\bf r}_T|{\bf r}_T)$, and therefore asymptotically the infinite space Green function $G_0({\bf r}_T|{\bf r}_T)$, is independent of $k$ in the case of non compact exploration. This is checked numerically in Fig. \ref{fig:HTT}. Identifying in Eq. (\ref{scalingt}) $A_k=G_0({\bf r}_T|{\bf r}_T)/k$, which is finite in the case of non compact exploration, we finally obtain:
 \begin{equation}
 \Tm_{\rr_T}(r) \sim N\langle k\rangle\left(
\frac{A}{k} - B r^{d_w-d_f} \right).
\label{scalingk2}
\end{equation}
As in the compact case this expression is valid for $r$ large, and becomes hypothetical for $r$ small. It  reveals that in the case of non compact exploration, the MFPT is independent of $r$ for $r$ large, and scales as the inverse connectivity of the target. This behavior is in strong contrast with the case of compact exploration.

\section{Summary of the results and discussion}
\begin{figure}[htb!]
\includegraphics[width =0.9 \linewidth,clip]{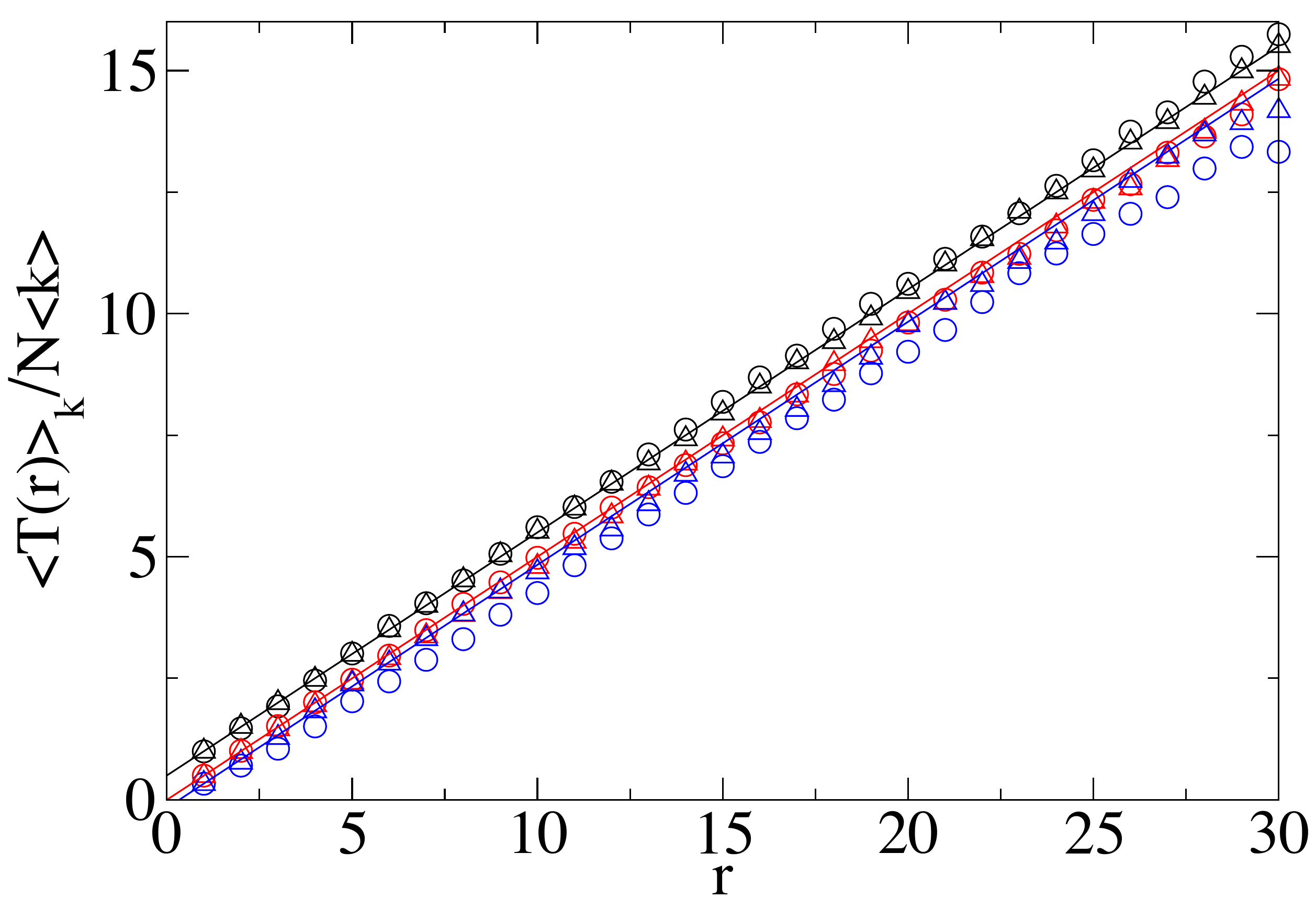} 
\caption{Mean First Passage Time ($\langle\Tm_{\rr_T}(r)\rangle_k$) for critical Erdos-Renyi networks, as a function of the source-target distance $r$, for a various target connectivity $k$. Circles and triangles stand for simulation results, for two sizes of the network (1000 and 2000 nodes), straight lines stand for the zero-constant formula ($\langle k \rangle = 2$) of Eq. (\ref{scaling4}).
}
\label{fig:Erdos}
\end{figure}
Finally our central result can be summarized as follows, where the case of marginal exploration ($d_w=d_f$) has been obtained along the same line :
\begin{equation}\label{scaling4}
\frac{\Tm_{\rr_T}(r)}{N \langle k \rangle} \! \sim \! \left \{ \begin{array}{ll}
\displaystyle \frac{1}{k}+\frac{1}{\langle k \rangle} \left (r^{d_w-d_f} - 1 \right )& \!\! \textnormal{ if } d_w > d_f\\[4mm]
\displaystyle \frac{1}{k}+ B \ln (r) & \!\! \textnormal{ if } d_w = d_f\\[4mm]
\displaystyle \frac{A}{k} - B r^{d_w-d_f} & \!\! \textnormal{ if } d_w < d_f
\end{array}. \right .
\end{equation}
This expression is very general and shows the respective impact of distance and connectivity on the MFPT. In particular the MFPT is {\it fully explicitly determined} in the compact case. The positive constants $A$ and $B$ depend on the network in the case of non compact exploration. We comment that in both cases the target connectivity $k$ plays an important role at short distances $r$. However for large source-target distances $r$, the $k$-dependence is damped out in the compact case, while it remains important in the non compact case. The $r$-dependence is found to be important in the compact case  and largely irrelevant in the non compact case in agreement with previous results \cite{Condamin:2007zl}.  The question raised in introduction can therefore be answered as follows : in the non compact case connected targets are found the fastest almost independently of their distance, while in the compact case close targets are found the fastest almost  independently of their connectivity.
\begin{figure}[htb!]
\includegraphics[width =0.9\linewidth,clip]{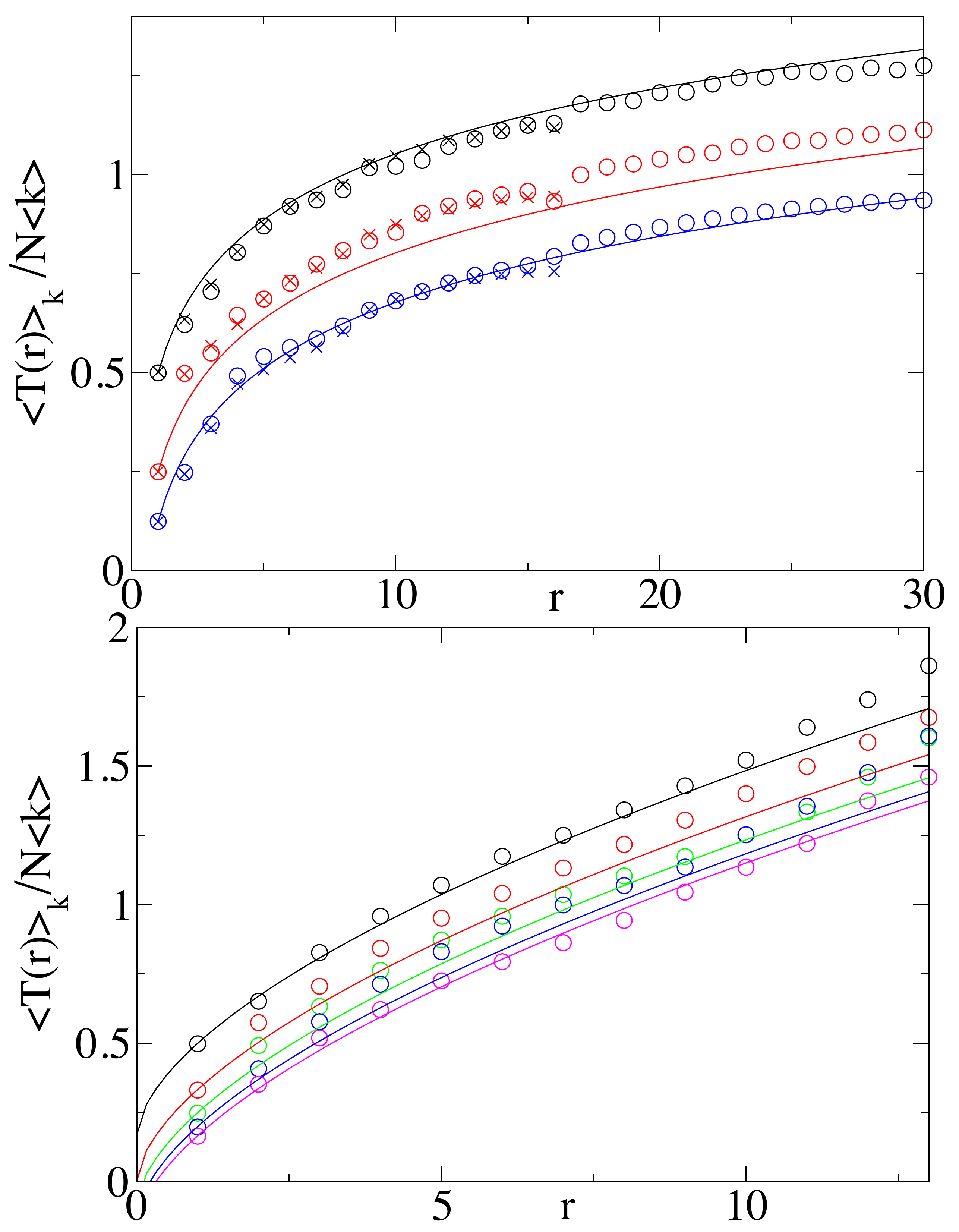}
\caption{Mean First Passage Time ($\langle\Tm_{\rr_T}(r)\rangle_k$) as a function of the source-target distance $r$, for a various target connectivity $k$. {\it Up :} $(2,2)$-flowers ($d_w=d_f$). Circles and triangles stand for simulation results, for two sizes of the network (generations 4 and 5), straight lines stand for the formula $1/k+B \ln(r)$ of eq. (\ref{scaling4}), with $B=0.24$. {\it Down:} random $(2,2)$-flowers ($d_w = 2.5$ and $d_f = 1.9$). Circles stand for simulations results, straight lines for $1/k+1/\langle k \rangle(r^{d_w-d_f})$ ($\langle k \rangle = 3$) of equation (\ref{scaling4}).
}
\label{Flowers}
\label{rflowersk}
\end{figure}
\begin{figure}[htb!]
\includegraphics[width =0.9\linewidth,clip]{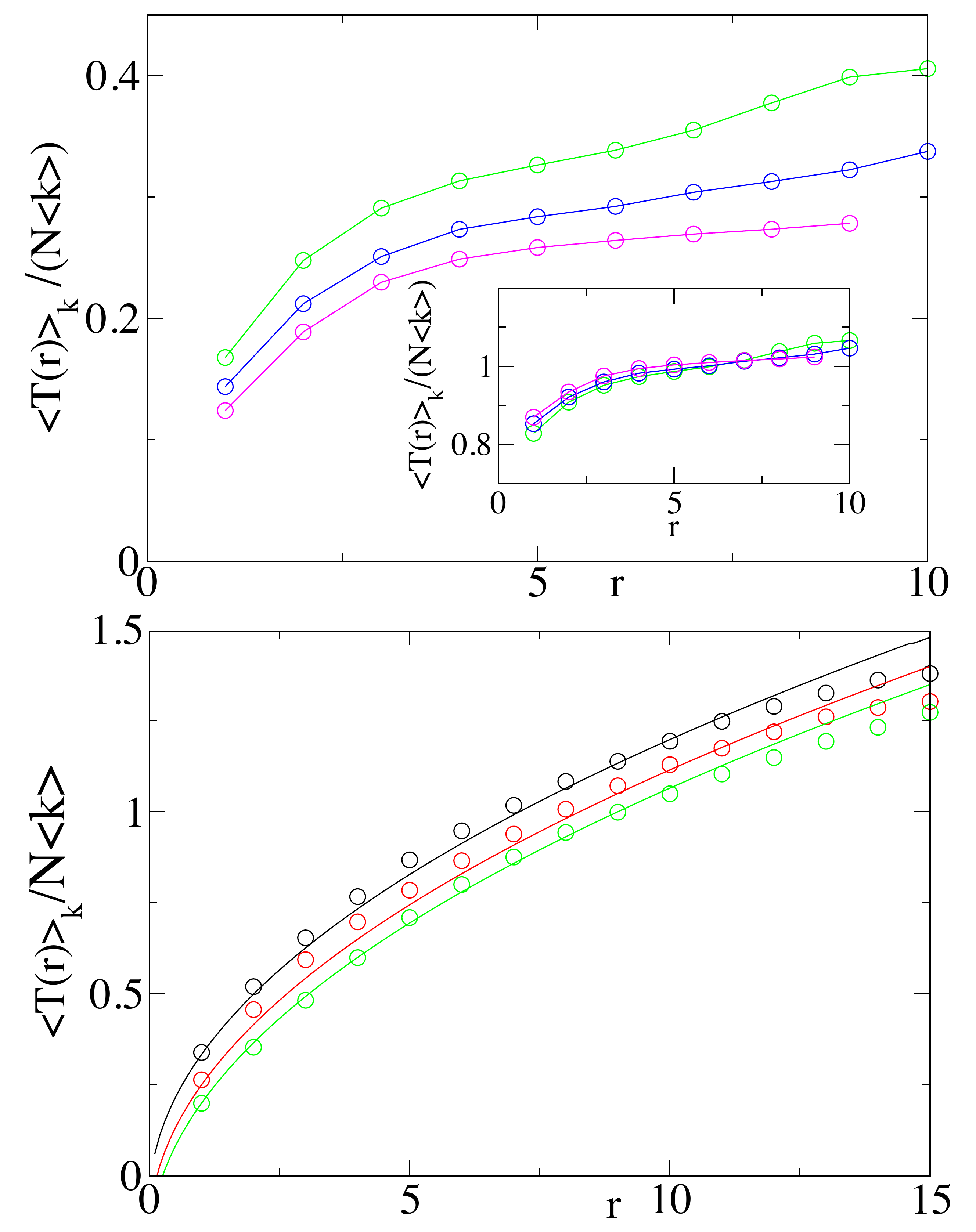}
\caption{Mean First Passage Time ($\langle\Tm_{\rr_T}(r)\rangle_k$) as a function of the source-target distance $r$, for a various target connectivity $k$. {\it Up}: non compact Kozma network of size $X=400$, $\alpha = 1.0$. The insight shows a translation along the $y$ axis of $A/k$ with $A=2.04$ according to equation (\ref{scalingb}). As predicted, this quantity does not depend on $k$. {\it Down}: compact Kozma network of size $X=50$, $\alpha = 2.5$. The expected scaling is in $r^{0.5}$: circles stand for simulation results, straight lines stand for $1/k+1/\langle k \rangle ( r^{0.5}-1)$ ($\langle k \rangle = 2.5$) of equation (\ref{scaling4}).}
\label{fig:Kozma}
\end{figure}
\begin{figure}[htb!]
\includegraphics[width =0.9\linewidth,clip]{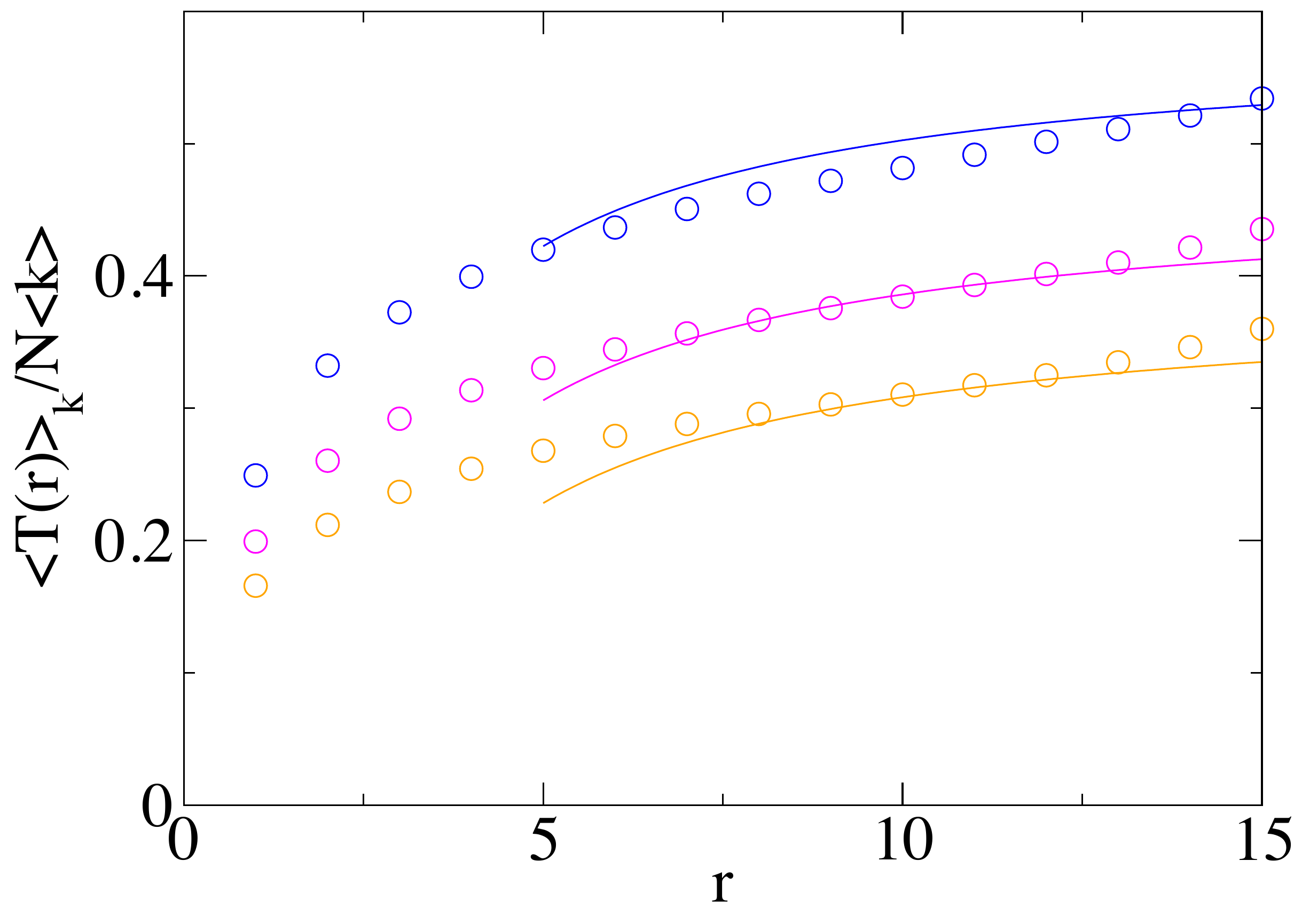}
\caption{Mean First Passage Time ($\langle\Tm_{\rr_T}(r)\rangle_k$) as a function of the source-target distance $r$, for a various target connectivity $k$, on supercritical 3D percolation network ($p=0.8$). For this network, $d_w \simeq 2$ and $d_f = 3$, the exploration is non compact. Circles stand for simulations results, straight lines for a fit by $A/k+B r^{d_w-d_f}$, with $A \simeq 2.33$ and $B \simeq 0.8$.}
\label{fig:percol3D}
\end{figure}

 We can conclude that for  self-similar networks with short range  degree correlations, the main criterion that governs the behavior of $\Tm$ is the type (compact or non compact) of the random walk. In particular the existence of loops is irrelevant.  Further comments are in order. (i) As stressed above, Eq. (\ref{scaling4}) is derived in the large $r$ regime. Its applicability to the small $r$ regime relies on the assumption that the scaling form of the propagator (\ref{scalingk}) holds for all values of $r$, which is not always satisfied for real networks.  In particular when degree correlations exist the relation $B=1/\langle k\rangle$ obtained in the compact case gives only a rough estimate, and the result of  Eq. (\ref{scaling4}) is valid only for $r$ larger than the correlation length. 
(ii) Our results can be extended to the case of non self-similar networks, still under  the assumption that degree correlations are negligible. Following the method developed above,  one can infer that
\begin{equation}\label{scalingb}
\Tm_{\rr_T}(r) \sim 
N\langle k\rangle\left(A/k+g(r) \right) 
\end{equation}
where $g$ does not depend on $k$ and satifies $g(r\to\infty)=C$ in the transient case, and $g(r\to\infty)=\infty$  in the recurrent case. 
The relative impact of connectivity and distance is therefore qualitatively the same as in the case of self-similar networks discussed above.
(iii) Incidentally, our results straightforwardly yield the $k$ dependence of the MFPT averaged over all source points (global MFPT). We find in the large $N$ limit :
\begin{equation}\label{scalinggmfpt}
\{\Tm_{\rr_T} \} \sim \left \{ \begin{array}{ll}
CN^{d_w/d_f} &  \textnormal{ if } d_w > d_f\\
CN\ln N &  \textnormal{ if } d_w = d_f\\
C N/k &  \textnormal{ if } d_w < d_f
\end{array} \right .
\end{equation}
which complements previous results obtained in \cite{Tejedor:2009vn}. This expression, along with Eq. (\ref{global}), yields as a by-product the large $N$ asymptotics of $H(\rr_T|\rr_T)$:
\begin{equation}\label{scalinghtt}
H(\rr_T|\rr_T) \sim \left \{ \begin{array}{ll}
CkN^{d_w/d_f-1} &  \textnormal{ if } d_w > d_f\\
Ck\ln N &  \textnormal{ if } d_w = d_f\\
C  &  \textnormal{ if } d_w < d_f
\end{array}. \right .
\end{equation}
This $k$-dependence of $H(\rr_T|\rr_T)$ is checked numerically in Fig. \ref{fig:HTT} and directly validates the $k$--dependence of the global MFPT .

\section{Numerical simulations}
We have checked our main result  (\ref{scaling4}) on various examples of networks, corresponding to compact or non compact random walks as detailed below. We stress  that the zero constant formula obtained in the compact case is in good agreement with numerical simulations in all the examples that we have considered.

{\it Erdos-Renyi networks} -- Erdos-Renyi networks can be defined as a  percolation cluster on a complete graph: for every pair of nodes  $(i,j)$, a link exists with probability $p$. The network is then defined as the largest cluster. We considered clusters at the percolation transition obtained for  $p = 1/N$, for which the estimated $d_f$ is $1,9-2,0$\cite{Song:2007}.  We computed numerically $d_w \simeq 2,9$, which shows that exploration is compact. Numerical results of Fig. \ref{fig:Erdos} are in very good agreement with the scaling (\ref{scaling4}).

{\it (u,v)--flowers} -- These networks are constructed recursively as described in \cite{Rozenfeld:2007}: at each step, every link is substituted by two paths of length $u$ and $v$. We extended this definition to $(u,v,w)$--flowers, for which a third path is added. For those networks, $d_w-d_f = -\ln(1/u+1/v+1/w)/\ln(u)$ (if $u \le v \le w$). Fig. \ref{rflowersk} shows a very good agreement of numerical simulations with equation (\ref{scaling4}), despite the small size of the networks. 

{\it Random flowers} -- These networks are constructed recursively  as described in \cite{Tian}: at each step, every link is substituted by two paths of length $u$ and $v$. 
$d_f$ and $d_w$ are determined numerically for those networks; in our example $(2,2)$-random flowers are compact networks ($d_w-d_f \simeq 0,6$). Fig. \ref{rflowersk} shows a good agreement of numerical simulations with equation (\ref{scaling4}). 

{\it Networks of Kozma et al.} -- These networks, defined in \cite{Kozma:2005}, are simple euclidian lattices in which long range links ("short-cuts") are added. A short-cut starts from each node  with  probability $p$, and leads to a node at a distance $r$ where $r$ is distributed according to a power law of index $\alpha$.   We consider here a $1D$ euclidian lattice. Exploration is then compact for $\alpha > 2$ and non-compact for $\alpha < 2$.  Again, Fig. \ref{fig:Kozma} shows a very good agreement of numerical simulations with equation (\ref{scaling4}).

{\it Percolation clusters} -- We consider percolation clusters in the case of bond percolation in $3D$  cubic lattices.  The critical probability is $p_c = 0.2488...$  and one has $d_w=3.88...$ and $d_f=2.58...$ at criticality. If $p > p_c$, $d_f = 3$ (euclidian dimension) and $d_w = 2$. Fig. \ref{fig:percol3D} shows a good agreement of numerical simulations with equation (\ref{scaling4}).

\section{Conclusion}
To conclude, we have proposed a general theoretical framework  which  elucidates the connectivity and source-target distance dependence of the MFPT for random walks on networks. This approach leads to explicit solutions for self-similar networks and  highlights 
two strongly different behaviors depending on the  type -- compact or non compact -- of the random walk. In the case of non compact exploration, the MFPT is found to scale as the inverse connectivity of the target,  and to be widely independent of the source-target distance. On the contrary, in the compact case the MFPT is controlled by the source-target distance, and we find thatunexpectedly the target connectivity is \textit{irrelevant} for remote targets. The question raised in introduction can therefore be answered as follows : in the non compact case connected targets are found the fastest almost independently of their position, while in the compact case close targets are found the fastest almost  independently of their connectivity.
Last, we stress that following \cite{Condamin:2008}, this explicit determination of MFPTs  can be straightforwardly generalized to obtain other relevant first-passage observables, such as splitting probabilities or occupation times.

\bibliographystyle{vincent}
\begin{acknowledgements}
V. T. wishes to thank Dr. Hakim Lakmini for useful discussions.
\end{acknowledgements}


\end{document}